\documentclass[lettersize,journal]{IEEEtran}
\usepackage{setspace,amsmath,latexsym,amssymb,epsfig,amsfonts}
\usepackage{amsmath}
\usepackage{amsthm}
\usepackage{url}
\usepackage{float}
\usepackage{graphicx}
\usepackage{mdframed}
\usepackage{epstopdf}
\usepackage{tikz}
\usepackage{pgfplots}
\usepackage{flushend}
\usepackage{verbatim}
\usepackage{algorithm}
\usepackage{algorithmic}
\usepackage{bm}
\usepackage[colorlinks]{hyperref}

\usepackage{balance}
\begin{document}

\title{Iterative Sparse Asymptotic Minimum Variance Based Channel Estimation in Fluid Antenna System}

\author{Zhen~Chen,~\IEEEmembership{Senior Member,~IEEE,}
            Jianqing~Li,~\emph{Senior Member},~\emph{IEEE},
            Xiu~Yin~Zhang,~\IEEEmembership{Fellow,~IEEE,}\\          
            Kai-Kit~Wong,~\IEEEmembership{Fellow,~IEEE},
            Chan-Byoung Chae,~\IEEEmembership{Fellow,~IEEE}, and
            Yangyang Zhang
\vspace{-7mm}

\thanks{This work has been supported in part by the National Natural Science Foundation of China under Grant 62371197,  in part by the Natural Science Foundation of Guangdong Province under Grant 2024A1515011172, in part by the Open Research Fund of National Mobile Communications Research Laboratory, Southeast University (No. 2019D06).}
\thanks{The work of K.-K. Wong is supported by the Engineering and Physical Sciences Research Council (EPSRC) under Grant EP/W026813/1.}
\thanks{The work of C. B. Chae is supported by the Institute for Information and Communication Technology Planning and Evaluation (IITP)/NRF grant funded by the Ministry of Science and ICT (MSIT), South Korea, under Grant RS-2024-00428780 and 2022R1A5A1027646.}

\thanks{S. Xu is with the School of Computer Science and Engineering, Macau University of Science and Technology, China-Macau (e-mail: sqxu1985@gmail.com).}
\thanks{Z. Chen is with the Institute of Microelectronics, University of Macau, China-Macau (e-mail: chenz.scut@gmail.com).}
\thanks{X. Zhang is with the School of Electronic and Information Engineering, South China University of Technology, Guangzhou, China (e-mail: xiuyinzhang@foxmail.com).}
\thanks{J. Li is with  the School of Computer Science and Engineering, Macau University of Science and Technology, Macau (e-mail: jqli@must.edu.mo).}
\thanks{K. K. Wong is with the Department of Electronic and Electrical Engineering, University College London, London WC1E 7JE, United Kingdom, and he is also affiliated with Yonsei Frontier Laboratory, Yonsei University, Seoul, 03722, Republic of Korea (e-mail: kai-kit.wong@ucl.ac.uk).}
\thanks{C. B. Chae is with School of Integrated Technology, Yonsei University, Seoul, 03722, Republic of Korea (e-mail: cbchae@yonsei.ac.kr).}
\thanks{Y. Zhang is with Kuang-Chi Science Limited, Hong Kong SAR, China.}}

\markboth{}%
{Shell \MakeLowercase{\textit{et al.}}: A Sample Article Using IEEEtran.cls for IEEE Journals}


\maketitle

\begin{abstract}
With fluid antenna system (FAS) gradually establishing itself as a possible enabling technology for next generation wireless communications, channel estimation for FAS has become a pressing issue. Existing methodologies however face limitations in noise suppression. To overcome this, in this paper, we propose a maximum likelihood (ML)-based channel estimation approach tailored for FAS systems, designed to mitigate noise interference and enhance estimation accuracy. By capitalizing on the inherent sparsity of wireless channels, we integrate an ML-based iterative tomographic algorithm to systematically reduce noise perturbations during the channel estimation process. Furthermore, the proposed approach leverages spatial correlation within the FAS channel to optimize estimation accuracy and spectral efficiency. Simulation results confirm the efficacy of the proposed method, demonstrating superior channel estimation accuracy and robustness compared to existing benchmark techniques.
\end{abstract}

\begin{IEEEkeywords}
Fluid antenna system (FAS), channel estimation, maximum likelihood (ML), iterative approach.
\end{IEEEkeywords}

\section{Introduction}
\IEEEPARstart{W}{ith} the sixth-generation (6G) wireless networks setting to transform our daily life, fluid antenna systems (FAS) have emerged as a possible enabling technology, offering unprecedented adaptability and performance enhancements over traditional antenna systems (TAS)
\cite{Wong-2022fcn,New-2024tut,Lu-2025}. Unlike TAS, FAS introduces a dynamic paradigm wherein antennas can adjust both their shape and position in real time to optimize key performance metrics such as gain, radiation patterns, and etc. for ultimate spatial diversity \cite{9131873,Wong-2021fas}. This capability is driven by advanced designs, including liquid-based antennas \cite{Huang-2021,shen2024design,Shamim-2025}, reconfigurable radio-frequency (RF) pixels \cite{zhang2024pixel,Shen-2024}, mechanically movable antenna structures \cite{basbug2017design}, and meta-material antennas \cite{Liu-2025arxiv}, which are reshaping the landscape of modern wireless communication. In particular, the designs in \cite{zhang2024pixel,Shen-2024,Liu-2025arxiv} facilitate fast adjustments of radiation points in no time, enabling real-time antenna reconfiguration.

\subsection{Prior Work}
Prior to FAS, attempts to incorporate the degree of freedom (DoF) of reconfigurable antennas into the physical layer could be found in 2004 by Cetiner {\em et al.}~in \cite{1367557} in which changes in operating frequency and polarization were considered in conjunction with space-time coding. Reconfiguration of antenna spacings in multiple-input multiple-output (MIMO) channels was later envisaged in \cite{4200712}. This was as far as the integration of reconfigurable antennas with the physical layer went before FAS. It was not until \cite{9131873,Wong-2021fas} when the scope of reconfigurable antennas was expanded to include all forms of position and shape flexibility in antennas and applied in the physical layer for spatial diversity gains and multiplexing benefits.

To harness the potential of FAS, extensive research has been conducted. For example, existing researches have attempted to improve the channel modelling of spatial correlation over the ports of FAS, e.g., \cite{202256432,Khammassi2023,Espinosa-2024}. Specifically, accurate modelling such as the one in \cite{Khammassi2023} normally leads to mathematical intractability. As a consequence, the authors in \cite{Espinosa-2024} presented a methodology that carries out the analysis using the simplified model in \cite{202256432} and then adopts a block model to obtain accurate performance analysis for FAS channels. Diversity order for a point-to-point FAS Rayleigh fading channel was analyzed in \cite{New2023fluid}. Recently, \cite{Alvim2023on} studied the performance of FAS under $\alpha$-$\mu$ fading channels. Copulas were found to be a powerful tool in the performance analysis of FAS channels \cite{Ghadi-2023,10319727,Ghadi-2024}. Most recently in \cite{new2023information}, the diversity-multiplexing trade-off for MIMO-FAS channels was also given. Joint antenna positioning (a.k.a.~port selection) and beamforming has also become a key problem for enhancing system capacity \cite{10416896,10473750,Xu-2024cap}.

In addition to these advancements, FAS provides a radically new way for multiple access. Specifically, the position reconfigurability of FAS allows a user terminal to exploit the spatial opportunity where the superposition of the interfering signals cancels. By doing so, the work in \cite{9650760} introduced a fast fluid antenna multiple access (FAMA) approach, which can support simultaneous communication of hundreds of users on the same physical channel, without precoding at the base station (BS) nor interference cancellation on the user side. Later in \cite{wong2023sFAMA}, a slow version of FAMA was proposed that changes its port if the channel state information (CSI) changes, but compromises the number of supportable users. Deep learning has also been shown to be effective in port selection for slow FAMA \cite{Waqar-2023,Eskandari-2024}. In \cite{10354059}, a mean-field game formulation was considered to optimize a slow FAMA network for greater energy efficiency. Moreover, \cite{Wong2024cuma} upgraded the slow FAMA scheme to consider analogue signal mixing from the selection of many `coherent' ports in order to serve a larger number of users. This scheme was understood to perform well in the presence of randomized reconfigurable intelligent surfaces (RISs) \cite{Wong-2024cuma-ris}. Recent efforts have also found channel coding being a useful tool to improve FAMA \cite{hong2024coded,hong2025Downlink}. Additionally, opportunistic scheduling was also considered to combine with FAMA in \cite{10078147,Waqar-2024ofama}.

Furthermore, the reconfigurability of FAS has spurred research into its integration with other emerging technologies, such as non-orthogonal multiple access (NOMA) \cite{10318134}, physical layer security \cite{Tang-2023,Xu-2024pls,Ghadi-2024dec}, RIS \cite{10539238} or fluid RIS \cite{salem2025first}, index modulation \cite{10794591,Yang-2024pim}, full-duplex communication \cite{103292959}, integrated sensing and communication (ISAC) \cite{Wang-2024isac,10705114,zhou2024fasisac}, simultaneous wireless information and power transfer \cite{10506795} and many more. These synergies further highlight the potential of FAS to revolutionize wireless communication by addressing challenges related to scalability and efficiency.

Despite the encouraging prospects of FAS, including high capacity, energy efficiency, and improved multiplexing, there remains a significant challenge. Its effectiveness heavily depends on the precise positioning of fluid antennas, which requires accurate and timely CSI. This accuracy is crucial for optimizing antenna placement and ensuring system performance. However, most research on FAS has assumed the availability of perfect CSI for all possible antenna positions within a given space. While this assumption simplifies theoretical analysis, perfect CSI is often impractical in real-world environments due to hardware limitations, noisy surroundings and the dynamic nature of wireless channels. Besides, the system must estimate CSI not only for fixed antenna positions but across the entire spatial region to which fluid antennas can access. As such, the CSI requirements of these tasks present a significant barrier to fully realizing the theoretical gains of FAS.

To estimate the CSI for FAS, a sequential linear minimum mean square error (MMSE)-based scheme was proposed \cite{9992289}. Compressed sensing techniques have also been investigated to exploit the spatial correlation of FAS channels \cite{10497534,10236898}. In addition, deep learning model was utilized to predict the CSI of unobserved antenna positions based on sampled locations, significantly reducing the overhead associated with estimation \cite{10299674,10615662,10495003}. Furthermore, low-sample-size sparse channel reconstruction techniques have proven effective by leveraging the inherent sparsity of FAS channels. These methods require fewer measurements, making them particularly advantageous in resource-limited environments \cite{10375559}. Moreover, kernel-based sampling and regression techniques, as proposed in the successive Bayesian reconstructor in \cite{10807122}, can enable efficient CSI acquisition. Specifically, Bayesian learning frameworks incorporate prior knowledge and therefore can further enhance the accuracy of sparse channel estimation \cite{9521836}.

\begin{figure*}
\centering
\includegraphics[width=1.6\columnwidth]{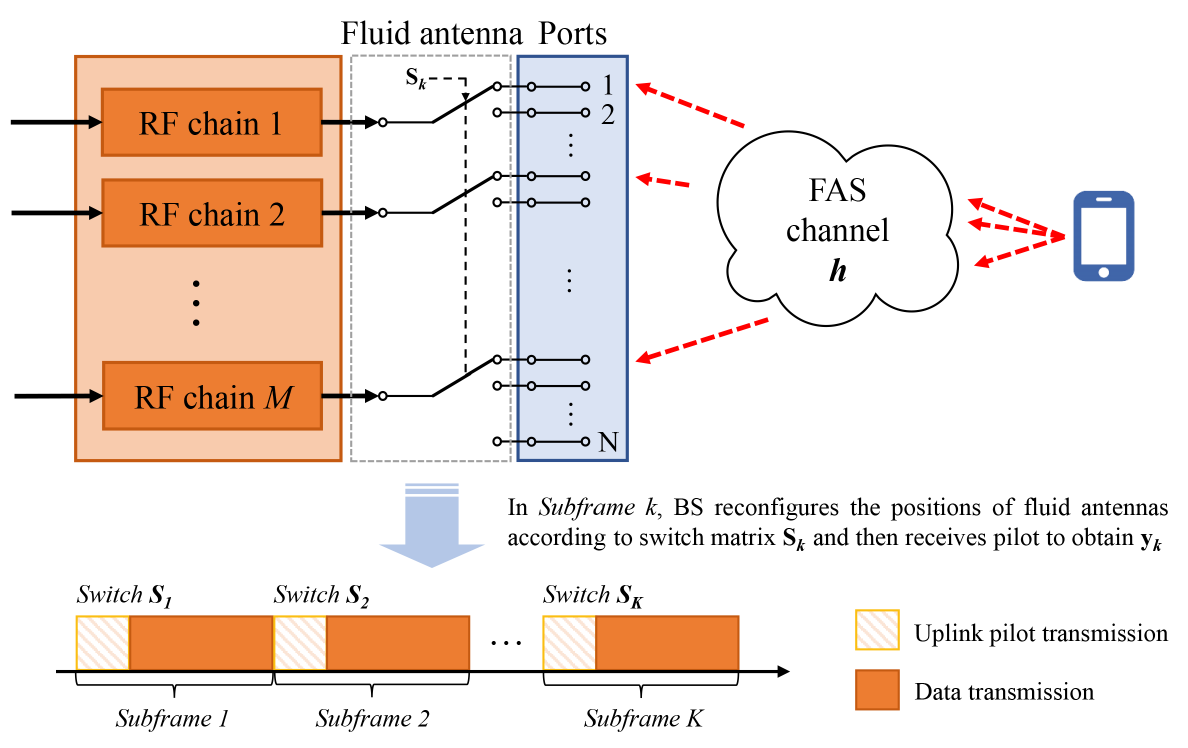}
\caption{An illustration of channel estimation for an FAS, where one $N$-port BS equipped with $M$ fluid antennas receives pilots.}\label{eq:fig01}
\end{figure*}

\subsection{Our Contributions}
Given the above literature, it is understood that perfect CSI reconstruction is impossible. This highlights the importance of spatial oversampling in FAS \cite{New-TWC2025}, regardless of the signal-to-noise ratio (SNR). Utilization of iterative spectral estimation methods has been shown to be beneficial in addressing some of these inaccuracies. Motivated by this, in this paper, we propose a new framework for FAS channel estimator (FAS-CHE) as a general solution to FAS channel estimation. In our proposal, the FAS channel and noise estimation problem is modeled as maximum likelihood (ML) estimation. Thus the expectation-maximization (EM) algorithm \cite{197789754} can be used to iteratively solve this problem and handle the noise suppression.

Specifically, our contributions are summarized as follows:
\begin{itemize}
\item To accommodate varying antenna configurations,  the idea of ML estimator is introduced into FAS channel estimation to track signal variations, thereby overcoming noise perturbations in the FAS channel estimation.
\item To balance between accuracy of the reconstructed channel and the number of estimated FAS channels, we develop a family of ML-based iterative tomographic algorithms, with improved accuracy and robustness, as well as the ability to work in the case of limited pilot signals.
\item For performance evaluation, we present comprehensive simulation results that demonstrate the practicality and efficiency of estimating and reconstructing the FAS channel, where an approximate solution to the noise estimator is also provided for different SNR conditions.
\end{itemize}

The remainder is organized as follows. Section \ref{sec:model} presents the system model for FAS. In Section \ref{sec:ce}, we then discuss the channel estimation problem while deriving important parameters, noise estimator. Simulation results are provided in Section \ref{sec:result}. Finally, the conclusion is drawn in Section \ref{sec:conclude}.

The following notations are adopted throughout this paper. Lower-case and upper-case boldface letters $\textbf{\emph{a}}$ and $\textbf{\emph{A}}$ denote a vector and a matrix, respectively; $\mathrm{tr}(\cdot)$ represents the trace operator; $\emph{Re}[\cdot]$ takes the real part of a complex variable; $\|\cdot\|_2$ represents the $L_2$ norm. Also, $(\cdot)^T$, $(\cdot)^H$ and $(\cdot)^*$ denote the transpose, conjugate transpose and conjugate, respectively.

\section{System Model and Existing Methods}\label{sec:model}
\subsection{System Model}
Our system consists of a transmitter equipped with an $N$-port FAS and a receiver with a fixed-position antenna. The FAS consists of  $M$ RF chains, with $M \ll N$. As illustrated in Fig.~\ref{eq:fig01}, each fluid antenna is connected to an RF chain for pilot signal reception, and its position can be switched among $N$ available port locations. Let $\emph{\textbf{h}}\in \mathbb{C}^N$ be the channel vectors corresponding to the $N$ ports, and $K$ be the number of transmit pilots during a coherence time frame. During each timeslot (or subframe), see Fig.~\ref{eq:fig01}, the fluid antennas adjust their positions to receive pilot signals. Specifically, for the first $K$ subframes, the $M$ RF chains can dynamically switch positions to capture pilot signals. The positions of the $M$ RF chains in timeslot $\emph{k}$ are represented by a binary indicator matrix $\textbf{\emph{S}}_{\emph{k}}\in\{0, 1\}^{M\times N}$, referred to as the switch matrix \cite{10807122}. We assume that $\textbf{\emph{S}}_{\emph{k}}$ is defined within an $A\times A$ square region. The matrix entries indicate connections: a value of 1 at position $(\emph{m, n})$ signifies that the $\emph{m}$-th RF chain is connected to the $\emph{n}$-th port, while a value of 0 indicates no connection.

With a fixed number of RF chains, only $M$ of $N$ ports can be selected at each slot. Since each RF chain connects to one port at a time, each row of $\textbf{\emph{S}}_\emph{k}$ contains only a single $1$, i.e.,
\begin{equation}\label{eq:2}
\|\textbf{\emph{S}}_{\emph{k}}(:,\emph{m})\|_2=1,~\forall \emph{m}\in\{1,\dots,M\}.
\end{equation}
Furthermore, two or more RF chains cannot be connected to the same port simultaneously, so the sum of each column of $\textbf{\emph{S}}_{\emph{k}}$ is either $0$ or $1$, i.e.,
\begin{equation}\label{eq:3}
\|\textbf{\emph{S}}_{\emph{k}}(\emph{n},:)\|\in\{0,1\},~\forall \emph{n}\in\{1,\dots,N\}.
\end{equation}
Depending on the actual design, one may prefer avoiding the coupling effect between adjacent active ports and choose to impose the following condition on $\emph{\textbf{S}}_{\emph{k}}$:
\begin{equation}\label{eq:4}
|(p_{i+1}-p_i)d|\geq\frac{\lambda}{2},~\forall p_i, p_{i+1}\in\Omega_{\emph{k}},
\end{equation}
where $\lambda$ is the carrier wavelength, $d = \frac{W\lambda}{N-1}$ is a port spacing with $W$ being the normalized size of the fluid antenna, $\Omega_{\emph{k}}$ represents the set of the selected ports at slot $\emph{k}$, while $p_i$ and $p_{i+1}$ denote the selected port indices.

The received signals at the BS in time slot $\emph{k}$ can be written in vector form as
\begin{equation}\label{eq:5}
{\bf{y}}_{\emph{k}} = \textbf{\emph{S}}_{\emph{k}}\textbf{\emph{h}}\emph{x}_{\emph{k}} + \bm{\varepsilon}_{\emph{k}},
\end{equation}
where $\emph{x}_\emph{k}$ denotes the pilot transmitted by the transmitter and $\bm{\varepsilon}_\emph{k}\sim\mathcal{CN}(\textbf{0}_M,\sigma\textbf{\emph{I}}_M)$ is the additive white Gaussian noise (AWGN) at the $M$ selected ports. Without loss of generality, we assume that $\emph{x}_\emph{k} = 1$ for all $\emph{k}\in\{1,\dots, K\}$. Considering the total $K$ slots for pilot transmission, we have
\begin{equation}\label{eq:6}
{\bf{y}} = \textbf{\emph{S}}\textbf{\emph{h}}  + \bm{\varepsilon},
\end{equation}
in which ${\bf{y}}=[{\bf{y}}_1^T,{\bf{y}}_2^T,\dots,{\bf{y}}_K^T]^T$, ${\textbf{\emph{S}}}=[\textbf{\emph{S}}_1^T,\textbf{\emph{S}}_2^T,\dots,\textbf{\emph{S}}_K^T]^T$ and the noise term ${\bm{\varepsilon}}=[{\bm{\varepsilon}}_1^T,{\bm{\varepsilon}}_2^T,\dots,{\bm{\varepsilon}}_K^T]^T$ with $\mathbb{E}(\bm{\varepsilon}\bm{\varepsilon}^H)=\sigma\textbf{\emph{I}}_{KM}$. At the receiver side, the FAS selects the $M$ ports that are then connected to the RF chains for data transmission during subsequent subframes within the coherence time.

\subsection{Channel Model}
We consider a narrowband spatially sparse clustered channel model with $N_c$ clusters each containing $N_r$ rays. Under this model, the FAS channel can be modeled as
\begin{equation}\label{eq:7}
{\textbf{\emph{h}}} = \sqrt{\frac{N}{N_cN_r}}\sum\limits_{i=1}^{N_c}\sum\limits_{j=1}^{N_r}\alpha_{i,j}\textbf{\emph{a}}(\theta_{i,j}),
\end{equation}
where $\alpha_{i,j}$ and $\theta_{i,j}$ represent the complex path gain and the incident angle associated with the $j$-th ray in the $i$-th cluster, respectively. Also, the steering vector $\textbf{\emph{a}}(\theta_{i,j})$ is given by
\begin{equation}\label{eq:8}
\textbf{\emph{a}}(\theta_{\emph{i,j}})= \frac{1}{\sqrt{N}}\left[1, e^{-j\frac{2\pi}{\lambda}d\cos(\theta_{\emph{i,j}})},\dots,e^{-j\frac{2\pi}{\lambda}(N-1)d\cos(\theta_{\emph{i,j}})}\right]^T.
\end{equation}

Our objective is to reconstruct the $N$-dimensional channel $\textbf{\emph{h}}$ from the $KM$-dimensional noisy observation {\bf{y}} with $KM\ll N$ described in (\ref{eq:6}). We outline our proposed method next.

\subsection{Proposed Method}
In this section, we develop a ML-based recursive approach to jointly estimate the FAS channel and noise variance (i.e., $\sigma$) based on the asymptotically minimum variance criterion. ML-based recursive techniques are based on iterative evaluation of the pseudo-spectrum profile (PSP) of the unknown vector $\textbf{\emph{h}}$, starting from an initial estimate, typically obtained from covariance matrix reconstruction. We first assume  that $\textbf{\emph{x}}$ and $\bm{\varepsilon}$ are independent of each other, and the covariance matrix $\textbf{\emph{R}}$ at the $i$-th iteration can be expressed as
\begin{equation}\label{eq:08}
\textbf{\emph{R}}^{(i+1)} = \textbf{\emph{S}}{\bf{\Gamma}}^{(i)}\textbf{\emph{S}}^H + \sigma^{(i)}\textbf{\emph{I}}_N,
\end{equation}
where the covariance matrix $\textbf{\emph{R}}^{(i+1)}$ is traditionally estimated by $\textbf{\emph{R}} =\frac{{\bf{y}}{\bf{y}}^H}{N}$. When the signals are uncorrelated, the expected value $\mathbb{E}(\textbf{\emph{h}}\textbf{\emph{h}}^H)=\bm{\Gamma}$ is diagonal, where $\bm{\Gamma} =  \mathrm{diag}(\bm{\gamma}) $ and $\bm{\gamma} \triangleq [\gamma_1,\dots,\gamma_M ]$. However, when the $M$ signals are partially correlated, $\bm{\gamma}$ becomes non-diagonal and remains invertible. The formulation in (\ref{eq:08}) is widely employed in signal processing to model low-dimensional signals influenced by circular complex Gaussian (CCG) white noise. This framework is suitable for the channel modeling considered in this study.

With independent and identically distributed (i.i.d.)~observations, $\{{\bf{y}}_\emph{k}\}_{\emph{k}=1}^K$ follow complex elliptically symmetric (CES)  distributed data snapshots, with a probability density function (pdf) expressed as
\begin{equation}\label{eq:9}
p({\bf{y}}_\emph{k}) = |{\textbf{\emph{R}}}|^{-1}\emph{g}\left({\bf{y}}_\emph{k}^H{\textbf{\emph{R}}}^{-1}{\bf{y}}_\emph{k}\right),
\end{equation}
where ${\textbf{\emph{R}}}$ is an $N\times N$ Hermitian positive definite scattering matrix, $\emph{g}(\cdot)$ is non-negative, satisfying $\delta_{M,\emph{g}}\triangleq\int_0^{\infty}t^{M-1}\emph{g}(t)dt<\infty$. Since the observations ${\bf{y}}_\emph{k}$ are CES distributed, this enables efficient processing of CCG. Note that the CCG distribution is obtained by setting $\emph{g}(t)=e^{-t}$.

We consider the stochastic negative log-likelihood function corresponding to (\ref{eq:9}) as our cost function, i.e.,
\begin{equation}\label{eq:10}
\mathcal{L}(\bm{\gamma}) = K\log(|\textbf{\emph{R}}|)-\sum\limits_{\emph{k}=1}^K(\emph{g}\left({\bf{y}}_\emph{k}^H{\textbf{\emph{R}}}^{-1}{\bf{y}}_\emph{k})\right).
\end{equation}
It is found that for a CES distribution dependent on the density generator function $\emph{g}(\cdot)$, and the functional (\ref{eq:10}) depends nonlinearly on $\bm{\gamma}$ and $\sigma$ embedded in an overparameterized covariance matrix model
$\textbf{\emph{R}}=\sum_{\emph{m}=1}^M\gamma_{\emph{m}}\textbf{\emph{s}}_{\emph{m}}\textbf{\emph{s}}_{\emph{m}}^H+\sigma\textbf{\emph{I}}$. Minimization of (\ref{eq:10}) with respect to (w.r.t.) $\bm{\gamma}$ is however cumbersome. By applying the Woodbury matrix identity \cite{19891989}, the covariance matrix of the interference and noise can be defined as
\begin{equation}\label{eq:10-1}
\bm{Q}_{\emph{m}} \triangleq \textbf{\emph{R}} - \gamma_{\emph{m}}\textbf{\emph{s}}_{\emph{m}}\textbf{\emph{s}}_{\emph{m}}^H,~\emph{m} = 1,\dots, M.
\end{equation}
Applying the matrix inversion lemma to (\ref{eq:10-1}) yields
\begin{equation}\label{eq:11}
\textbf{\emph{R}}^{-1} = \textbf{\emph{Q}}_{\emph{m}}^{-1}-\gamma_\emph{m}\zeta_\emph{m}\textbf{\emph{w}}_\emph{m}\textbf{\emph{w}}_\emph{m}^H,~\emph{m}=1,\dots,M,
\end{equation}
where $\textbf{\emph{w}}_{\emph{m}} \triangleq \textbf{\emph{Q}}_{\emph{m}}^{-1}\textbf{\emph{s}}_\emph{m}$ and
\begin{equation}\label{eq:11-1}
\zeta_{\emph{m}}\triangleq \left(1 + \gamma_{\emph{m}}\textbf{\emph{s}}_{\emph{m}}^H\textbf{\emph{Q}}_\emph{m}^{-1}\textbf{\emph{s}}_{\emph{m}}\right)^{-1}.
\end{equation}
Then it follows that
\begin{equation}\label{eq:12}
{\bf{y}}_\emph{k}^H\textbf{\emph{Q}}_{\emph{m}}^{-1}{\bf{y}}_\emph{k}={\bf{y}}_\emph{k}^H\textbf{\emph{R}}^{-1}{\bf{y}}_\emph{k}+\gamma_\emph{m}\zeta_\emph{m}|\textbf{\emph{w}}_\emph{m}^H{\bf{y}}_\emph{k}|^2.
\end{equation}

Due to the well-known algebraic identity $\mathrm{det}(\textbf{\emph{I}} + \textbf{\emph{CD}}) = \mathrm{det}(\textbf{\emph{I}} + \textbf{\emph{DC}})$, we obtain
\begin{align}
\log(\mathrm{det}(\textbf{\emph{R}}))& = \log(\mathrm{det}(\textbf{\emph{Q}}_\emph{m} + \gamma_\emph{m}\textbf{\emph{s}}_\emph{m}\textbf{\emph{s}}_\emph{m}^H))\notag\\
& = \log\left(\mathrm{det}(\textbf{\emph{Q}}_\emph{m})(1 + \gamma_\emph{m}\textbf{\emph{s}}_\emph{m}^H\textbf{\emph{Q}}_\emph{m}^{-1}\textbf{\emph{s}}_\emph{m})\right)\notag\\
& = \log\left(\mathrm{det}(\textbf{\emph{Q}}_\emph{m})\right) - \log\left({\bf{\zeta}}_\emph{m}\right).\label{eq:13}
\end{align}
Substituting (\ref{eq:12}) and (\ref{eq:13}) into (\ref{eq:10}) results in
\begin{multline}\label{eq:14}
\mathcal{L}(\bm{\gamma})= K\log\left(\mathrm{det}(\textbf{\emph{Q}}_{\emph{m}})\right) - K\log\left(\mathrm{det}(\zeta_{\emph{m}})\right)\\
-\sum\limits_{\emph{k}=1}^K\log\left(\emph{g}({\bf{y}}_\emph{k}^H\textbf{\emph{Q}}_{\emph{m}}^{-1}{\bf{y}}_\emph{k}-\gamma_{\emph{m}}\zeta_\emph{m}|\textbf{\emph{w}}_\emph{m}^H{\bf{y}}_\emph{k}|^2)\right).
\end{multline}
The objective function has now been decomposed into nondependent $\gamma_\emph{m}$ terms and dependent $\gamma_\emph{m}$ terms. Therefore, (\ref{eq:10}) reduces to the following function that depends only on $\gamma_\emph{m}$, which gives rise to a simple optimization problem:
\begin{multline}\label{eq:15}
\mathcal{L}(\gamma_\emph{m}) = -K\log(\zeta_\emph{m})\\
-\sum\limits_{\emph{k}=1}^K\log\left(\emph{g}({\bf{y}}_\emph{k}^H\textbf{\emph{Q}}_{\emph{m}}^{-1}{\bf{y}}_\emph{k}-\gamma_\emph{m}\zeta_\emph{m}|\textbf{\emph{w}}_\emph{m}^H{\bf{y}}_\emph{k}|^2)\right).
\end{multline}

To find the values of $\{\gamma_\emph{m}\}_{\emph{m}=1}^M$ that minimize (\ref{eq:15}), we proceed in the standard way by finding the first-order derivative of (\ref{eq:15}) w.r.t.~$\gamma_\emph{m}$. Using the fact that $\gamma_{\emph{m}}(\textbf{\emph{s}}_\emph{m}^H\textbf{\emph{Q}}_\emph{m}^{-1}\textbf{\emph{s}}_\emph{m})\zeta_\emph{m}^2-\zeta_\emph{m}=-\zeta_\emph{m}^2$ and (\ref{eq:12}), the first-order derivative of (\ref{eq:15}) can be found as
\begin{equation}\label{eq:16}
\begin{aligned}
      \mathcal{L}^\prime(\gamma_{\emph{m}}) &= K\zeta_\emph{m}(\textbf{\emph{s}}_\emph{m}^H\textbf{\emph{Q}}_\emph{m}^{-1}\textbf{\emph{s}}_\emph{m})
      +\left(\gamma_{\emph{m}}\zeta_\emph{m}^2(\textbf{\emph{s}}_\emph{m}^H\textbf{\emph{Q}}_\emph{m}^{-1}\textbf{\emph{s}}_\emph{m})-\zeta_{\emph{m}}\right)\\
      &\quad-\sum\limits_{\emph{k}=1}^K\varphi\left({\bf{y}}_\emph{k}^H\textbf{\emph{Q}}_{\emph{m}}^{-1}{\bf{y}}_\emph{k}-\gamma_{\emph{m}}\zeta_\emph{m}|\textbf{\emph{w}}_\emph{m}^H{\bf{y}}_\emph{k}|^2\right)|\textbf{\emph{w}}_\emph{m}^H{\bf{y}}_\emph{k}|^2\\
      &=
      K\zeta_\emph{m}(\textbf{\emph{s}}_\emph{m}^H\textbf{\emph{Q}}_\emph{m}^{-1}\textbf{\emph{s}}_\emph{m})\\
      &\quad- K\zeta_\emph{m}^2\textbf{\emph{w}}_\emph{m}^H\left(\frac{1}{K}\sum\limits_{\emph{k}=1}^K\varphi\left({\bf{y}}_\emph{k}^H\textbf{\emph{R}}^{-1}{\bf{y}}_\emph{k}\right){\bf{y}}_\emph{k}{\bf{y}}_\emph{k}^H\right)\textbf{\emph{w}}_\emph{m}^H
\end{aligned}
\end{equation}
According to (\ref{eq:16}), the $\mathcal{L}^\prime(\gamma_{\emph{m}})$ can be rearranged as
\begin{equation}\label{eq:17}
\mathcal{L}^\prime(\gamma_{\emph{m}})= K\zeta_{\emph{m}}(\textbf{\emph{s}}_\emph{m}^H\textbf{\emph{Q}}_\emph{m}^{-1}\textbf{\emph{s}}_\emph{m}) + K\zeta_{\emph{m}}^2\textbf{\emph{s}}_\emph{m}^H\textbf{\emph{Q}}_\emph{m}^{-1}\textbf{\emph{R}}_{K}\textbf{\emph{Q}}_\emph{m}^{-1}\textbf{\emph{s}}_\emph{m},
\end{equation}
where
\begin{equation}\label{eq:18}
\begin{aligned}
\textbf{\emph{R}}_{K}=\frac{1}{K}\sum_{\emph{k}=1}^K\kappa_\emph{k}{\bf{y}}_\emph{k}{\bf{y}}_\emph{k}^H
\end{aligned}
\end{equation}
with
\begin{equation}\label{eq:180}
\begin{aligned}
    \kappa_{\emph{k}}=\varphi({\bf{y}}_\emph{k}^H\textbf{\emph{R}}^{-1}{\bf{y}}_\emph{k})
\end{aligned}
\end{equation}
that satisfies Maronna's conditions  $\varphi(t)\triangleq-\frac{1}{\emph{g}(t)}\frac{d\emph{g}(t)}{dt}$ in \cite{19764432}. The bracketed sum is replaced by the ML  $M$-estimate of $\textbf{\emph{R}}$, denoted as $\textbf{\emph{R}}_K$. This estimate is defined as the solution to (\ref{eq:18}), assuming prior knowledge. When $\textbf{\emph{y}}_\emph{k}$ is CCG-distributed and $\varphi(t) = 1$, the estimate simplifies to $\hat{\textbf{\emph{R}}}_K\triangleq\frac{1}{K}\sum_{\emph{k}=1}^K{\bf{y}}_\emph{k}{\bf{y}}_\emph{k}^H $. Next, we derive the expression for $\gamma_{\emph{m}}$, which corresponds to the solution when (\ref{eq:17}) is set to $0$. By incorporating $\zeta_{\emph{m}}$ from (\ref{eq:11-1}), we further obtain
\begin{equation}\label{eq:19}
  \hat{\gamma}_{\emph{m}}=\frac{\textbf{\emph{s}}_\emph{m}^H\textbf{\emph{Q}}_\emph{m}^{-1}(\textbf{\emph{R}}_K-\textbf{\emph{Q}}_\emph{m})\textbf{\emph{Q}}_\emph{m}^{-1}\textbf{\emph{s}}_\emph{m}}{(\textbf{\emph{s}}_\emph{m}^H\textbf{\emph{Q}}_\emph{m}^{-1}\textbf{\emph{s}}_\emph{m})^2},
\end{equation}
for $\emph{m} = 1,\dots, M$, where $\textbf{\emph{R}}_K$ is given by (\ref{eq:18}). The second-order derivative of (\ref{eq:17}) gives
\begin{multline}\label{eq:20}
  \mathcal{L}^{\prime\prime}(\gamma_{\emph{m}})=K\zeta_\emph{m}^2\left(\textbf{\emph{s}}_\emph{m}^H\textbf{\emph{Q}}_\emph{m}^{-1}\textbf{\emph{s}}_\emph{m}\right)\\
\times \left(2\zeta_\emph{m}(\textbf{\emph{w}}_\emph{m}^H\textbf{\emph{R}}_K\textbf{\emph{w}}_\emph{m}) - \textbf{\emph{s}}_\emph{m}^H\textbf{\emph{Q}}_\emph{m}^{-1}\textbf{\emph{s}}_\emph{m}\right)
\end{multline}
and therefore, we obtain after replacing $\gamma_{\emph{m}}$ with $\hat{\gamma}_{\emph{m}}$ in (\ref{eq:19}) that
\begin{equation}\label{eq:21}
\begin{aligned}
  \mathcal{L}^{\prime\prime}(\gamma_{\emph{m}})= \frac{K(\textbf{\emph{s}}_\emph{m}^H\textbf{\emph{Q}}_\emph{m}^{-1}\textbf{\emph{s}}_\emph{m})^4}{(\textbf{\emph{s}}_\emph{m}^H\textbf{\emph{Q}}_\emph{m}^{-1}\textbf{\emph{R}}_K\textbf{\emph{Q}}_\emph{m}^{-1}\textbf{\emph{s}}_\emph{m})^2}>0,
\end{aligned}
\end{equation}
which then proves that $\hat{\gamma}_{\emph{m}}$ in (\ref{eq:19}) is the unique minimizer of (\ref{eq:18}). Note that $\hat{\bm{\gamma}} = (\hat{\gamma}_1,\dots, \hat{\gamma}_M )$ is strictly positive if $\textbf{\emph{s}}_\emph{m}^H\textbf{\emph{Q}}_\emph{m}^{-1}\textbf{\emph{R}}_K\textbf{\emph{Q}}_\emph{m}^{-1}\textbf{\emph{s}}_\emph{m}>\textbf{\emph{s}}_\emph{m}^H\textbf{\emph{Q}}_\emph{m}^{-1}\textbf{\emph{s}}_\emph{m}$. By using (\ref{eq:11}), we have
\begin{equation}\label{eq:22}
\textbf{\emph{s}}_\emph{m}^H\textbf{\emph{Q}}_\emph{m}^{-1}\textbf{\emph{s}}_\emph{m} = \left(1+\gamma_{\emph{m}}\textbf{\emph{s}}_\emph{m}^H\textbf{\emph{Q}}_\emph{m}^{-1}\textbf{\emph{s}}_\emph{m}\right)\left(\textbf{\emph{s}}_\emph{m}^H\textbf{\emph{R}}^{-1}\textbf{\emph{s}}_\emph{m}\right)
\end{equation}
and
\begin{multline}\label{eq:23}
\textbf{\emph{s}}_\emph{m}^H\textbf{\emph{Q}}_\emph{m}^{-1}\textbf{\emph{R}}_K\textbf{\emph{Q}}_\emph{m}^{-1}\textbf{\emph{s}}_\emph{m} = \\ \left(1+\gamma_{\emph{m}}\textbf{\emph{s}}_\emph{m}^H\textbf{\emph{Q}}_\emph{m}^{-1}\textbf{\emph{s}}_\emph{m}\right)^2
\left(\textbf{\emph{s}}_\emph{m}^H\textbf{\emph{R}}^{-1}\textbf{\emph{R}}_K\textbf{\emph{R}}^{-1}\textbf{\emph{s}}_\emph{m}\right).
\end{multline}

Substituting (\ref{eq:22}) and (\ref{eq:23}) into (\ref{eq:19}), we obtain
\begin{equation}\label{eq:24}
\begin{aligned}
   \hat{\gamma}_{\emph{m}} &\hspace{.5mm}= \frac{\textbf{\emph{s}}_\emph{m}^H\textbf{\emph{R}}^{-1}(\textbf{\emph{R}}_K-\textbf{\emph{R}})\textbf{\emph{R}}^{-1}\textbf{\emph{s}}_\emph{m}}{(\textbf{\emph{s}}_\emph{m}^H\textbf{\emph{R}}^{-1}\textbf{\emph{s}}_\emph{m})^2}\\
   &\qquad\quad-\frac{1}{\left(1+\gamma_{\emph{m}}\textbf{\emph{s}}_\emph{m}^H\textbf{\emph{Q}}_\emph{m}^{-1}\textbf{\emph{s}}_\emph{m}\right)\left(\textbf{\emph{s}}_\emph{m}^H\textbf{\emph{R}}^{-1}\textbf{\emph{s}}_\emph{m}\right)}\\
   &\stackrel{(a)}{=}\frac{\textbf{\emph{s}}_\emph{m}^H\textbf{\emph{R}}^{-1}\textbf{\emph{R}}_K\textbf{\emph{R}}^{-1}\textbf{\emph{s}}_\emph{m}}{(\textbf{\emph{s}}_\emph{m}^H\textbf{\emph{R}}^{-1}\textbf{\emph{s}}_\emph{m})^2}+\gamma_{\emph{m}}
   -\frac{1}{\textbf{\emph{s}}_\emph{m}^H\textbf{\emph{R}}^{-1}\textbf{\emph{s}}_\emph{m}},
\end{aligned}
\end{equation}
where (\emph{a}) is due to the fact that
\begin{equation}\label{eq:24-1}
1+\gamma_{\emph{m}}\textbf{\emph{s}}_\emph{m}^H\textbf{\emph{Q}}_\emph{m}^{-1}\textbf{\emph{s}}_\emph{m}=1+\gamma_{\emph{m}}\chi_{\emph{m}}^{-1}\textbf{\emph{s}}_\emph{m}^H\textbf{\emph{R}}_\emph{m}^{-1}\textbf{\emph{s}}_\emph{m}
\end{equation}
leads to $\chi_{\emph{m}}=1-\gamma_{\emph{m}}\textbf{\emph{s}}_\emph{m}^H\textbf{\emph{R}}_\emph{m}^{-1}\textbf{\emph{s}}_\emph{m}$.
Differentiating (\ref{eq:10}) w.r.t.~variance $\sigma$ and equating to $0$, we get
\begin{equation}\label{eq:25}
\begin{aligned}
   K&\mathrm{tr}(\textbf{\emph{R}}^{-1})-K\mathrm{tr}\left(\textbf{\emph{R}}^{-2}\left(\frac{1}{K}\sum\limits_{\emph{k}=1}^K\varphi\left({\bf{y}}_\emph{k}^H\textbf{\emph{R}}^{-1}{\bf{y}}_\emph{k}\right){\bf{y}}_\emph{k}{\bf{y}}_\emph{k}^H\right)\right)\\
   &=\mathrm{tr}\left(\textbf{\emph{R}}^{-1}\right) - \mathrm{tr}\left(\textbf{\emph{R}}^{-2}\textbf{\emph{R}}_K\right)\\
   &=\mathrm{tr}\left(\textbf{\emph{R}}^{-2}\left(\textbf{\emph{R}} - \textbf{\emph{R}}_K\right)\right)\\
   &=\mathrm{tr}\left(\textbf{\emph{R}}^{-2}\left(\hat{\textbf{\emph{R}}} + \sigma\textbf{\emph{I}} - \textbf{\emph{R}}_K\right)\right)\\
   &=\sigma\mathrm{tr}\left(\textbf{\emph{R}}^{-2}\right)+\mathrm{tr}\left(\textbf{\emph{R}}^{-2}(\hat{\textbf{\emph{R}}} - \textbf{\emph{R}}_K)\right)\\
   &= 0.
\end{aligned}
\end{equation}
As such, the solution of (\ref{eq:25}) w.r.t.~variance $\sigma$ is given as
\begin{equation}\label{eq:26}
\hat{\sigma} = \frac{\mathrm{tr}\left(\textbf{\emph{R}}^{-1}(\textbf{\emph{R}}_K-\hat{\textbf{\emph{R}}})\textbf{\emph{R}}^{-1}\right)}{\mathrm{tr}(\textbf{\emph{R}}^{-2})},
\end{equation}
which after replacing $\textbf{\emph{R}}-\sigma\textbf{\emph{I}}$ by $\hat{\textbf{\emph{R}}}$ in the above, gives
\begin{align}
\hat{\sigma} &= \frac{\mathrm{tr}\left(\textbf{\emph{R}}^{-1}(\textbf{\emph{R}}_K-\textbf{\emph{R}})\textbf{\emph{R}}^{-1}\right)}{\mathrm{tr}\left(\textbf{\emph{R}}^{-2}\right)}+\sigma\notag\\
&=\frac{\mathrm{tr}\left(\textbf{\emph{R}}^{-2}\textbf{\emph{R}}_K\right)}{\mathrm{tr}\left(\textbf{\emph{R}}^{-2}\right)}+\sigma-\frac{\mathrm{tr}(\textbf{\emph{R}}^{-1})}{\mathrm{tr}\left(\textbf{\emph{R}}^{-2}\right)}.\label{eq:27}
\end{align}

Since $\hat{\gamma}_{\emph{m}}$ and $\hat{\sigma}$ depend on the knowledge of $\gamma_{\emph{m}}$, $\sigma$ and $\textbf{\emph{R}}$, this can  be realized in an iterative manner by
\begin{multline}\label{eq:28}
\hat{\gamma}_\emph{m}^{(i+1)} = \frac{\textbf{\emph{s}}_\emph{m}^H\bm{\Upsilon}_K^{(i)}\textbf{\emph{s}}_\emph{m}}{\left(\textbf{\emph{s}}_\emph{m}^H\left(\frac{1}{\textbf{\emph{R}}}\right)^{(i)}\textbf{\emph{s}}_\emph{m}\right)^2}+\gamma_\emph{m}^{(i)}
   -\frac{1}{\textbf{\emph{s}}_\emph{m}^H\left(\frac{1}{\textbf{\emph{R}}}\right)^{(i)}\textbf{\emph{s}}_\emph{m}},\\
\emph{m} = 1,\dots,M,
\end{multline}
and
\begin{equation}\label{eq:29}
\hat{\sigma}^{(i+1)} = \frac{\mathrm{tr}\left(\bm{\Upsilon}_K^{(i)}\right)+\hat{\sigma}^{(i)}\mathrm{tr}\left(\left(\frac{1}{\textbf{\emph{R}}^2}\right)^{(i)}\right)-\mathrm{tr}\left(\left(\frac{1}{\textbf{\emph{R}}}\right)^{(i)}\right)}{\mathrm{tr}\left(\textbf{\emph{R}}^{-2}\right)^{(i)}},
\end{equation}
where $\textbf{\emph{R}}^{(i)}$ and $\bm{\Upsilon}_K^{(i)}$ that are involved in the developed updating scheme are defined in terms of the estimated parameters at the $i$-th iteration as
\begin{align}
\textbf{\emph{R}}^{(i)} &= \sum\limits_{\emph{m}=1}^M\gamma_{\emph{m}}^{(i)}\textbf{\emph{s}}_\emph{m}\textbf{\emph{s}}_\emph{m}^H +\hat{\sigma}^{(i)}\textbf{\emph{I}},\label{eq:30}\\
\bm{\Upsilon}_K^{(i)} &\triangleq \left(\frac{1}{\textbf{\emph{R}}}\right)^{(i)}\textbf{\emph{R}}_K\left(\frac{1}{\textbf{\emph{R}}}\right)^{(i)}.\label{eq:31}
\end{align}

The estimates given by (\ref{eq:28}) and (\ref{eq:29}) might give irrational negative values due to the presence of the non-zero terms $\gamma_{\emph{m}}-\frac{1}{\textbf{\emph{s}}_\emph{m}^H\textbf{\emph{R}}^{-1}\textbf{\emph{s}}_\emph{m}}$ and $\sigma-\frac{\mathrm{tr}(\textbf{\emph{R}}^{-1})}{\mathrm{tr}(\textbf{\emph{R}}^{-2})}$. To resolve this problem, the non-negativity of the power estimation can be enforced by
\begin{multline}\label{eq:32}
\hat{\gamma}_\emph{m}^{(i+1)} = \\
\mathrm{max}\left\{0,\frac{\textbf{\emph{s}}_\emph{m}^H\bm{\Upsilon}^{(i)}\textbf{\emph{s}}_\emph{m}}{\left(\textbf{\emph{s}}_\emph{k}^H(\textbf{\emph{R}}^{-1})^{(i)}\textbf{\emph{s}}_\emph{m}\right)^2}+\gamma_\emph{m}^{(i)}
   -\frac{1}{\textbf{\emph{s}}_\emph{m}^H(\textbf{\emph{R}}^{-1})^{(i)}\textbf{\emph{s}}_\emph{m}}\right\},\\
\emph{m} = 1,\dots,M,
\end{multline}
and
\begin{multline}\label{eq:33}
\hat{\sigma}^{(i+1)} = \\
\mathrm{max}\left\{0,\frac{\mathrm{tr}\left(\bm{\Upsilon}_K^{(i)}\right)+\hat{\sigma}^{(i)}\mathrm{tr}\left(\left(\frac{1}{\textbf{\emph{R}}^2}\right)^{(i)}\right)-\mathrm{tr}\left(\left(\frac{1}{\textbf{\emph{R}}}\right)^{(i)}\right)}{\mathrm{tr}(\textbf{\emph{R}}^{-2})^{(i)}}\right\}.
\end{multline}

The algorithm requires iterations due to the need for knowledge of $\textbf{\emph{R}}$ and the values of $\{\hat{\gamma}_{\emph{m}}, \hat{\sigma}\}$ from the $i$-th iteration to update the formulas for $\{\hat{\gamma}_\emph{m}, \hat{\sigma}\}$ at the $(i+1)$-th iteration. To begin the process, the initialization of $\hat{\gamma}_{\emph{m}}^{(0)}$ can be set as
\begin{equation}\label{eq:34}
\hat{\gamma}_{\emph{m}}^{(0)} = \frac{\textbf{\emph{s}}_\emph{m}^H\textbf{\emph{R}}_K\textbf{\emph{s}}_\emph{m}}{\|\textbf{\emph{s}}_\emph{m}\|^4},~\emph{m} = 1,\dots,M.
\end{equation}
Also, the noise variance estimator can be initialized as
\begin{equation}\label{eq:35}
\hat{\sigma}^{(0)} = \frac{1}{MK}\sum\limits_{\emph{k}=1}^K\|{\bf{\textbf{y}}}\|_2^2.
\end{equation}
These approximate expressions are derived at differents level of SNR separately, compared to the unified expressions (\ref{eq:32}) and (\ref{eq:33}) regardless of SNR values or the number of sources. The overall proposed algorithm is given as Algorithm \ref{alg}.

\begin{algorithm}[!t]
\caption{Proposed Iterative Tomographic Algorithm}\label{alg}
\begin{algorithmic}[1]
        \STATE \textbf{Input}  $N$ available port locations; $M$ RF chains;  binary indicator matrix $\textbf{\emph{S}}_{\emph{k}}$; the observed vector ${\bf{y}};$
        \STATE \textbf{Initialize:}
        \STATE \quad Calculate the $\textbf{\emph{R}}_K^{(0)}$  according to (\ref{eq:18});
        \STATE \quad Calculate the $\gamma_{\emph{m}}^{(0)}$  according to (\ref{eq:34});
        \STATE \quad Calculate the $\sigma^{(0)}$  according to (\ref{eq:35});
        \STATE \textbf{while} $N_{\mathrm{max}}$ not reached \textbf{do}
		\STATE \quad $i=i+1$;
        \STATE \quad Update $\textbf{\emph{R}}^{(i+1)} = \textbf{\emph{S}}{\bf{\Gamma}}^{(i)}\textbf{\emph{S}}^H + \sigma^{(i)}\textbf{\emph{I}}_N$ according to (\ref{eq:11})
        \STATE \quad Update the weight $\kappa_{\emph{k}}$ according to (\ref{eq:180});
        \STATE \quad Update the $\textbf{\emph{R}}^{(i+1)}$ according to (\ref{eq:30});
        \STATE \quad Update the $\bm{\Upsilon}_K^{(i+1)}$ according to (\ref{eq:31});
        \STATE \quad \textbf{For} $\emph{m} = 1,...,M$ \textbf{do}
        \STATE \qquad Update the $\hat{\gamma}_\emph{m}^{(i+1)}$ according to (\ref{eq:32});
        \STATE \qquad Update the $\hat{\sigma}^{(i+1)}$ according to (\ref{eq:33});
        \STATE \quad \textbf{End}
		\STATE \textbf{end while}
        \STATE \textbf{Output}: $\gamma$, $\hat{\sigma}$
	\end{algorithmic}
\end{algorithm}

\section{Enhanced FAS Channel Estimation}\label{sec:ce}
As mentioned above, negative values that occur during the iterative estimates of (\ref{eq:24}) and (\ref{eq:27}) are avoided by (\ref{eq:32}) and (\ref{eq:33}), respectively. Based on them, the similarity between the solution operators for the iterative approach can be explained. Since the term $(\textbf{\emph{R}}^{(i)})^{-1}\textbf{\emph{s}}$ is common, $\gamma_\emph{m}^{(i)}$ and $\textbf{\emph{s}}_\emph{k}^H(\textbf{\emph{R}}^{-1})^{(i)}\textbf{\emph{s}}_\emph{m}$ can be raised to the powers $(0, 1, 0.5)$ and $(1, 0, 0.5)$ respectively. Thus, a general formula can be proposed by introducing a regularization parameter $\rho$ that can be used to balance between the ML-based methods. The solution operator becomes
\begin{equation}\label{eq:37}
\hat{\gamma}_\emph{m}^{(i+1)} = \frac{\textbf{\emph{s}}_\emph{m}^H\bm{\Upsilon}^{(i)}\textbf{\emph{s}}_\emph{m}}{\left(\textbf{\emph{s}}_\emph{k}^H(\textbf{\emph{R}}^{-1})^{(i)}\textbf{\emph{s}}_\emph{m}\right)^{2\rho}}\left(\gamma_\emph{m}^{(i)}\right)^{2(1-\rho)}.
\end{equation}

Considering Eq.(\ref{eq:37}), it is evident that when the parameter $\rho$ is set to 0.5, the proposed algorithm effectively blends the previous iteration power spectrum estimation with a beamforming technique specifically, the Capon method \cite{67582345}. The Capon beamformer, also known as the minimum variance distortionless response (MVDR) beamformer, aims to minimize the output power of the array while maintaining a distortionless response in the direction of the desired signal. This integration enables the iterative refinement of the preliminary 
PSP by enhancing the contributions of dominant scatterers, while concurrently suppressing spurious peaks arising from noise and multipath interference in the wireless fading channel. In particular, the strategy selectively reinforces the dominant scatterer due to its relatively stronger channel gain, leading to improved spectral resolution under high SNR conditions. However, to ensure robust detection in environments containing multiple scatterers, especially under the low SNR scenarios, the choice of $\rho$ becomes critical. By empirically determining an appropriate value for $\rho$, the enhanced FAS-CHE scheme can achieve a balance between refinement and generalization. Under such tuning, the enhanced FAS-CHE scheme converges to a solution that significantly reduces nondominant scatterers without unintentionally eliminating meaningful secondary scatterers within the propagation environment.

\begin{figure}
\centering
\includegraphics[width=0.95\columnwidth]{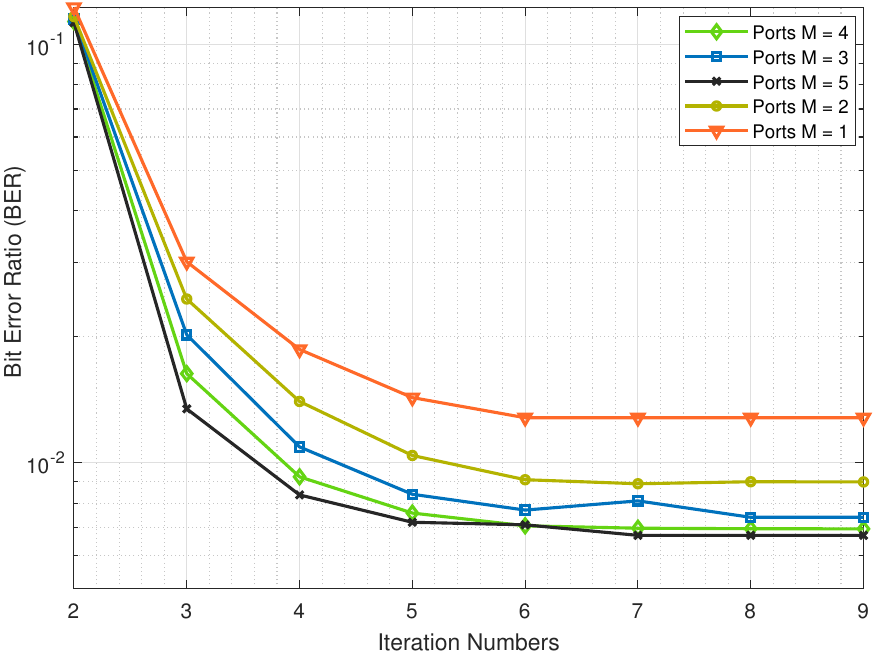}
\caption{Convergence performance for port selection based on the proposed FAS-CHE scheme against the  numbers of iteration.}\label{eq:fig07}
\end{figure}

\section{Simulation Results}\label{sec:result}
In this section, we assess the effectiveness of our proposed approach against the state-of-the-art methods. To conduct this evaluation, we define the receiver SNR based on the normalized transmit power that is given by SNR $=\mathbb{E}\left(\frac{\textbf{\emph{hh}}^H}{\sigma}\right)$. The default value $\sigma$ is set to different values to verify robustness. The performance is evaluated by the normalized mean square error (NMSE). We simulate both model-mismatched and model-matched scenarios for existing parametric estimators. These simulations are based on two channel models: the QuaDRiGa channel model from \cite{6758357} and the spatially-sparse clustered (SSC) channel model in (\ref{eq:7}), with channel parameters from 3GPP TR 38.901 \cite{10268998}. For the simulations, we consider five FAS channel estimators: SeCE \cite{9992289}, OMP-FAS \cite{10236898}, LS-FAS, Proposed FAS-CHE, and Enhanced FAS-CHE.

The results in Fig.~\ref{eq:fig07} examine the convergence performance for port selection based on the proposed FAS-CHE scheme against the  numbers of iteration. As observed, proposed FAS-CHE scheme with different port selection rapidly converges to a constant value after just a few iterations, regardless of the port selection number. This further demonstrating the efficiency and quick stabilization of our proposed FAS-CHE scheme. Additionally, we also find that the bit error ratio (BER) reduces with the number of ports increasing. These results have further verified the effectiveness of our proposed FAS-CHE scheme.

\begin{figure}
\centering
\includegraphics[width=0.9\columnwidth]{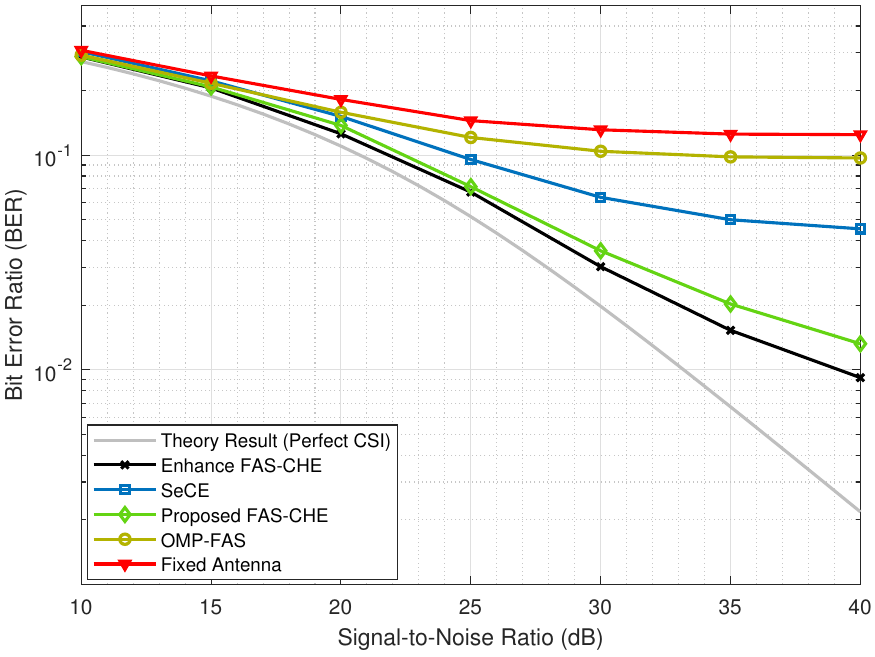}
\caption{The BER versus the receiver SNR under the QuaDRiGa model.}\label{eq:fig05}
\end{figure}

\begin{figure}
\centering
\includegraphics[width=0.9\columnwidth]{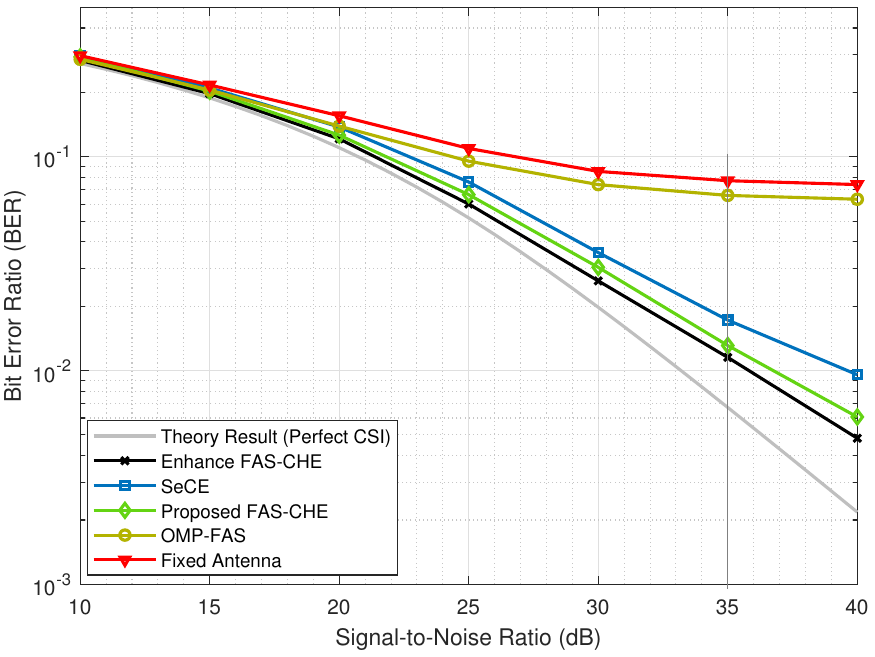}
\caption{The BER versus the receiver SNR under the SSC model in (\ref{eq:7}).}\label{eq:fig06}
\end{figure}

Next, Figs.~\ref{eq:fig05} and \ref{eq:fig06} compare the BER results of each algorithm with respect to the different level of SNR value, which follow the assumptions of QuaDRiGa model and SSC model in (\ref{eq:7}), respectively. As the receiver SNR value increases, the BERs of all estimators decrease rapidly. Particularly, the proposed FAS-CHE scheme achieves the highest estimation accuracy for both SSC and QuaDRiGa channels. The superior performance of the proposed FAS-CHE scheme can be attributed to its effective utilization of noise prior knowledge and the port selection. In contrast, existing methods do not leverage these prior knowledge. For OMP-FAS and LS-FAS schemes, the measured ports are selected randomly, meaning that the information provided by the measured channels may fail to capture the noise prior knowledge of the fading channel. While SeCE partially exploits channel correlation, it uses zero-order interpolation to estimate the unmeasured channels, neglecting the potential channel estimation errors. On the other hand, the proposed FAS-CHE scheme incorporates noise $\sigma$ prior information into the estimator, which reduces potential estimation errors across all channels simultaneously. Moreover, unlike OMP-FAS and LS-FAS, which are sensitive to model mismatches, the FAS-CHE scheme does not rely on a specific channel model, making it more robust to such mismatches. This demonstrates the superior performance and versatility of the proposed scheme.

\begin{figure}
\centering
\includegraphics[width=0.9\columnwidth]{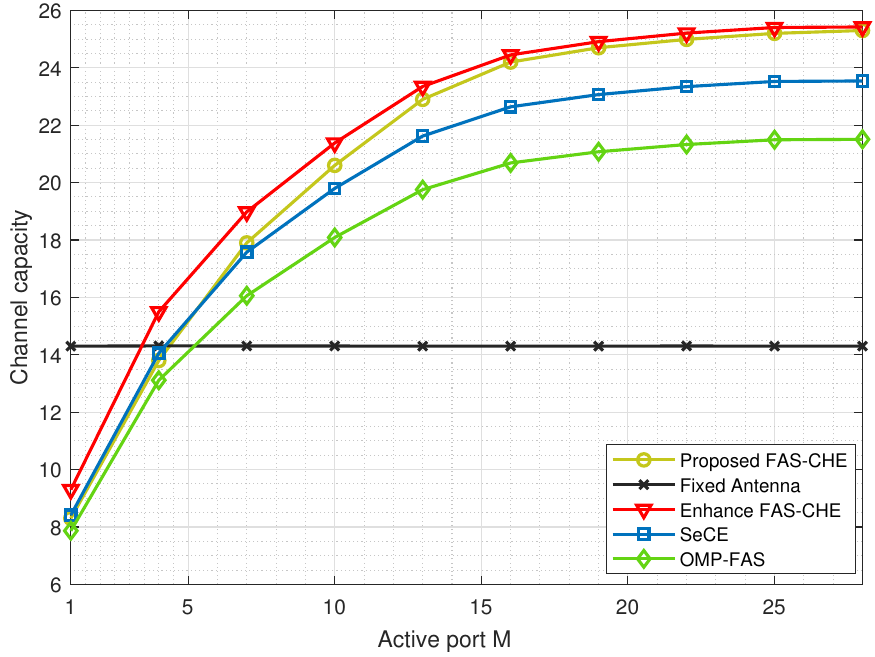}
\caption{The channel capacity versus the number of fluid antenna ports. }\label{eq:fig08}
\end{figure}

To satisfy the condition $M \ll N$, we restrict the horizontal axis to the range of $M = 2$ to $28$. As shown in Fig.~\ref{eq:fig08}, both of our proposed schemes achieve higher channel capacity across the entire fluid antenna port selection range compared to the benchmarks. However, when the number of fluid antenna ports is relatively  limited, the performance of our FAS-CHE scheme is initially worse than that of the enhanced FAS-CHE scheme. This discrepancy can be explained by the fact that, when the number of pilots is limited, although the FAS-CHE scheme is capable of identifying the optimal ports, the fixed regularization parameter $\rho$ prevents it from fully utilizing the fully advantage of port selection. Consequently, its performance is suboptimal in this scenario. On the other hand, as the number of fluid antenna ports increases beyond a certain threshold, our enhanced FAS-CHE scheme close to the proposed FAS-CHE scheme, highlighting its superior ability to leverage appropriate $\rho$ value and achieve better performance under smaller port numbers.

\begin{figure}
\centering
\includegraphics[width=0.9\columnwidth]{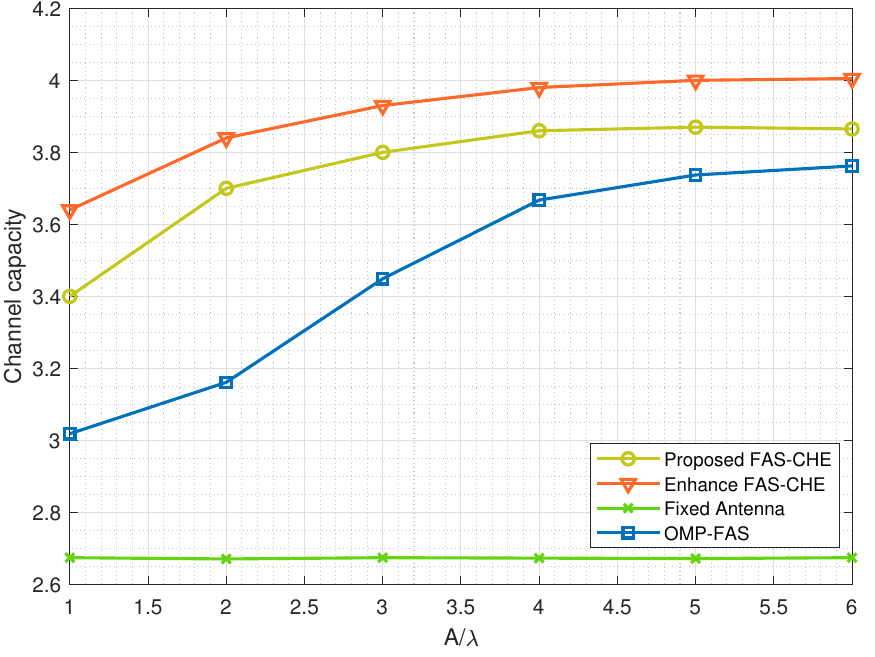}
\caption{The channel capacity versus $A/\lambda$ under the SSC model in (\ref{eq:7}).}\label{eq:fig09}
\end{figure}

Fig.~\ref{eq:fig09} illustrates the channel capacity of the proposed FAS-CHE scheme and baseline approaches as a function of the region size in timeslot $\emph{k}$, normalized by wavelength $A/\lambda$. It is observed that all FAS approaches, including both the FAS schemes and our proposed design, benefit from a larger antenna region. As the normalized region size $A/\lambda$ increases, particularly when it reaches 5, the curves for all FAS schemes converge to a near-stationary state, indicating that a limited FAS region can achieve the maximum channel capacity. Furthermore, the proposed FAS-CHE design consistently outperforms the two benchmark schemes across all simulation scenarios, with the fixed antenna scheme yielding the lowest performance. Specifically, when $A = 3.5\lambda$, the proposed FAS-CHE design achieves channel capacity of 3.93 dB and 3.82 dB, respectively. This demonstrates the significant advantage of the proposed approach in terms of communication performance.

\begin{figure}
\centering
\includegraphics[width=0.93\columnwidth]{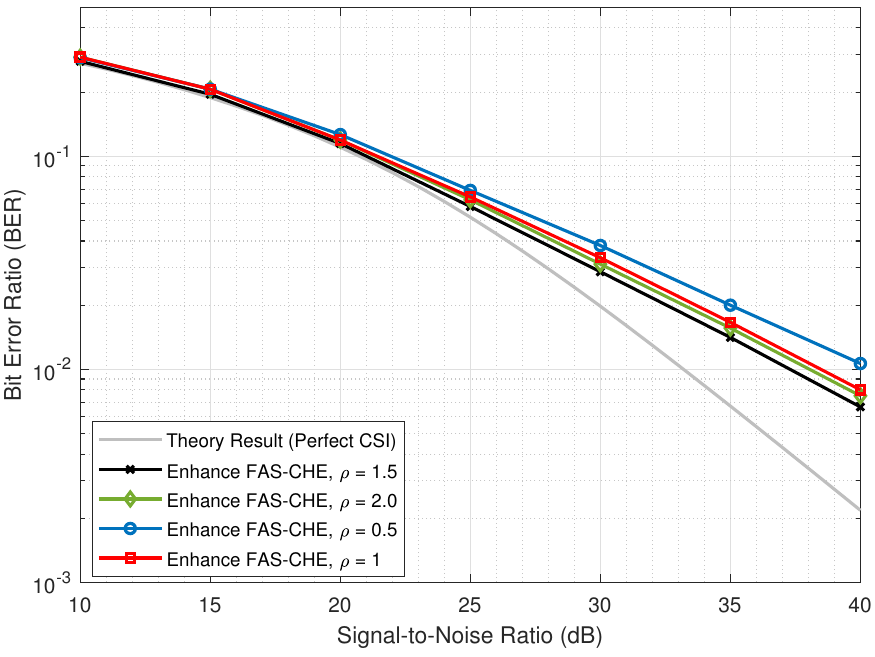}
\caption{The BER of the proposed FAS-CHE scheme for different $\rho$.}\label{eq:fig10}
\end{figure}

Finally, the performance of the enhanced  FAS-CHE scheme with different $\rho$ value is illustrated,  as well as the theoretical result under perfect CSI conditions. In Fig.~\ref{eq:fig10}, the theoretical result (perfect CSI), represented by the gray curve, which shows the lowest BER across all SNR values, indicating the optimal performance limit. As the SNR increases, the BER of all schemes decreases, which is expected as higher SNR leads to improved signal quality and reduced BER errors. Under the SNR $<$ 25 case, all parameters $\rho$ configurations closely follow the theoretical performance. However, when the the SNR level is greater than 25dB, all BER performance away from  the theoretical curve due to imperfect CSI scenario. Besides, it seems that the  BER performance exhibit slightly differences when the parameter is set to 1.5 and 2, respectively. Therefore, it suggests that higher $\rho$ values contribute to better BER performance, particularly for $\rho = 1.5$. The reason is that with the increase of the number of the fading channel, the enhanced  FAS-CHE scheme can obtain higher spatial diversity gain. Moreover, more array gain can be exploited by increasing the $\rho$ value, which consequently leads to an improvement in the BER performance of the system, as anticipated.

\section{Conclusion}\label{sec:conclude}
The paper developed a novel FAS channel estimation technique to address the challenges of noise perturbation and imperfect CSI in fluid antenna systems. The proposed ML-based iterative algorithm effectively mitigates noise interference, enhancing channel estimation accuracy and robustness, especially in scenarios with limited pilot signals. The study's findings indicate a significant improvement over existing methods, leveraging the sparsity of the FAS channel for better performance under varied SNR. Different existing literature that often struggling with noise suppression and hardware limitations, this work advances beyond these techniques by utilizing the EM algorithm to iteratively solve the channel and noise estimation problem, offering a more robust approach that effectively handles noise perturbations and ensures higher accuracy in FAS channel reconstruction.  Simulation results validate the effectiveness of the proposed FAS channel estimation strategies, demonstrating the feasibility and effectiveness of FAS systems.

\balance

\bibliographystyle{IEEEtran}


\vfill

\end{document}